%% file: arXiv.tex
\documentclass[11pt,letterpaper]{article}

\usepackage[utf8]{inputenc} 
\usepackage[margin=1in]{geometry}
\usepackage{graphicx}
\usepackage{algorithm}
\usepackage{algpseudocode}%
\usepackage[T1]{fontenc}
\usepackage{url}
\usepackage{ifthen}
\usepackage{enumitem}
\usepackage[cmex10]{amsmath}
\usepackage{amssymb}
\usepackage{cite}
\usepackage{amsthm}
\usepackage{textcomp}
\usepackage{siunitx}
\usepackage{color}
\usepackage[lofdepth,lotdepth]{subfig}
\usepackage[countmax]{subfloat}
\usepackage{cases}
\usepackage[font={small}]{caption}
\usepackage{tcolorbox}
\usepackage{fix-cm}
\usepackage{bbm}
\usepackage{pgfplots}

\makeatletter
\def\thm@space@setup{\thm@preskip=2pt
\thm@postskip=2pt \itshape}
\makeatother
\newtheoremstyle{newstyle}      
{} 
{} 
{\mdseries} 
{} 
{\bfseries} 
{.} 
{ } 
{} 

\theoremstyle{newstyle}

\newtheorem{theorem}{Theorem}

\theoremstyle{definition}

\theoremstyle{remark}
\newtheorem{remark}{Remark}

\setlist[description]{style=multiline}

\title{Coded State Machine\\ - Scaling State Machine Execution under Byzantine Faults}

\author{Songze~Li$^{*}$, Saeid Sahraei$^{*}$, Mingchao~Yu$^{*}$, Salman~Avestimehr$^{*}$\\ Sreeram Kannan$^{\dagger}$, and Pramod Viswanath$^{\ddagger}$\\
\\$^{*}$University of Southern California\\ $^{\dagger}$University of Washington\\$^{\ddagger}$University of Illinois at Urbana-Champaign\\Email: \{songzeli, ss\_805, mingchay\}@usc.edu,
avestimehr@ee.usc.edu\\ ksreeram@ee.washington.edu, pramodv@illinois.edu
}

\date{\vspace{-0ex}}
\begin{document}
\sloppy

\setlength{\abovedisplayskip}{1mm}
\setlength{\belowdisplayskip}{1mm}
\setlength{\abovecaptionskip}{1mm}
\setlength{\belowcaptionskip}{-6pt}

\maketitle

\input{abstract_new.tex}
\input{intro.tex}
\input{setting.tex}

\input{conventional.tex}
\input{results.tex}
\input{coded_state_machine.tex}

\input{throughput.tex}
\input{discussion.tex}

\newpage

\bibliographystyle{acm}
\bibliography{CSM.bib}

\appendix
\input{app.tex}

\end{document}

%% file: abstract_new.tex
\begin{abstract}
\normalsize
We introduce an information-theoretic framework, named Coded State Machine (CSM), to securely and efficiently execute multiple state machines on untrusted network nodes, some of which are Byzantine. The standard method of solving this problem is using State Machine Replication, which achieves high security at the cost of low efficiency. We propose CSM, which achieves the optimal linear scaling in storage efficiency, throughput, and security simultaneously with the size of the network. The storage efficiency is scaled via the design of Lagrange coded states and coded input commands that require the same storage size as their origins. The computational efficiency is scaled using a novel delegation algorithm, called INTERMIX, which is an information-theoretically verifiable matrix-vector multiplication algorithm of independent interest. Using INTERMIX, the network nodes securely delegate their coding operations to a single worker node, and a small group of randomly selected auditor nodes verify its correctness, so that computational efficiency can scale almost linearly with the network size, without compromising on security.
\end{abstract}

%% file: intro.tex
\section{Introduction}
\vspace{-3mm}

\newcommand{\bb}{}

\bb{A state machine is a program that executes a state transition function $(S(t+1),Y(t)) = f(S(t),X(t))$. In each round $t$, given an input command $X(t)$ and the current state $S(t)$, the machine uses $f$ to compute an output command $Y(t)$, and transitions its state to $S(t+1)$. In this paper, we consider the problem of securely implementing $K$ independent and identical state machines over a network of $N \geq K$ untrusted compute nodes, some of which are subject to Byzantine faults. Such model has a wide range of applications. For instance, multiple financial institutes manage their users' accounts over a data center comprised of commodity hardware, or a blockchain system maintains multiple ledgers (shards) of users' transactions over a peer-to-peer network.}

\bb{For such a system, there are three critical performance metrics: 1) security $\beta$, defined as the maximum number of malicious nodes that can be tolerated; 2) storage efficiency $\gamma$, defined as the total number of state machines that can be supported given that each node has a storage size of a single state; and 3) throughput $\lambda$, defined as the total number of commands processed per unit time. Information theoretically, all three metrics scale at most linearly with the network size $N$.}


\bb{However, simultaneous linear scaling of all three metrics has never been achieved in the literature. The canonical approach to deal with faulty nodes/processors is state machine replication (SMR) (see, e.g.,~\cite{alsberg1976principle,lamport1978implementation,schneider1990implementing,lamport1998part,castro1999practical}). The most classic one is {\em full replication}, where all the $K$ state machines are replicated at all the $N$ nodes. By applying the majority rule, full replication achieves a linearly-scaling security of $\beta_{\textup{full}} = \frac{N}{2}$, but a storage efficiency of only $\gamma_{\textup{full}}=1$, and a throughout of only $\lambda_{\textup{full}} = \frac{1}{c(f)}$, where $c(f)$ is the computational complexity of the state transition function $f$ (details in Section 3.1). An alternative approach is {\em partial replication}, where different state machines are replicated at a different disjoint subset of $q = N/K$ nodes. Partial replication improves the storage efficiency and throughout to $\gamma_{\textup{partial}} = K$ and $\lambda_{\textup{partial}} = \frac{K}{c(f)}$, respectively, which scale linearly with $N$ by letting $K$ scale linearly with $N$. However, since each state machine is only handled by a set of $q$ nodes at any time regardless of the allocation strategy \cite{kokoris2017omniledger,naor2003simple}, the security drops by $K$ times to $\beta_{\textup{partial}} = \frac{q}{2}$ once the adversary identifies this set and then corrupts it. A key theoretical question in this context is whether this tradeoff between efficiency and security is fundamental or can be circumvented by careful algorithm design.  }

\begin{table}[h]
\vspace{-0.5mm}
  \centering
\caption{Performance comparison of proposed CSM with state machine replication and information-theoretic limits in synchronous networks. Here $0 \leq \mu <\frac{1}{2}$ is some constant ($\mu = \frac{1}{3}$ is a concrete example). The state transition function $f$ is an arbitrary multivariate polynomial with constant degree $d$. $c(f)$ and $c(\textup{coding})$ are the computational complexities of $f$ and the coding operations per node in CSM respectively.}
 \label{table:compare}
 \vspace{-0.5mm}
  \begin{tabular}{| c | c | c | c | }
    \hline
    \rule{0pt}{11pt} & Security & Storage efficiency & Throughput  \\ \hline
    \rule{0pt}{11pt} Full Replication & $\frac{N}{2}$  & $1$ & $\frac{1}{c(f)}$  \\ \hline
    \rule{0pt}{11pt}  Partial Replication & $\frac{N}{2K}$ & $K$ & $\frac{K}{c(f)}$  \\ \hline
     \rule{0pt}{11pt} Information-Theoretic Limit & $\frac{N}{2}$  & $N$ & $\frac{N}{c(f)}$ \\ \hline
     \rule{0pt}{15pt} {\bf Coded State Machine} (CSM)& $\mu N$  & $(1-2\mu)N/d +\! 1 \!-\! 1/d$ & $\frac{(1-2\mu)N/d + 1 \!-\! 1/d}{c(f) + c(\textup{coding})}$
      \\\hline
  \end{tabular}
 \vspace{-2mm}
\end{table}

The main contribution of this paper is to demonstrate that there is {\em no} fundamental tradeoff between (storage and throughout) efficiency scaling and security scaling in state machine operation. In particular, we propose ``Coded State Machine'' (CSM), which {\em simultaneously} achieves linear scaling for both storage efficiency and security, for both synchronous and partially synchronous networks (see Table~\ref{table:compare}). CSM also achieves an almost linear scaling for throughput under synchronous networks through the development of an interactive matrix-vector multiplication verification protocol named INTERMIX (specifically, INTERMIX helps to reduce the coding complexity per node $c(\textup{coding})$ in Table~\ref{table:compare} to $O(\log^2 N \log \log N)$).



\begin{figure}[htbp]
  \centering
  \includegraphics[width=0.75\textwidth]{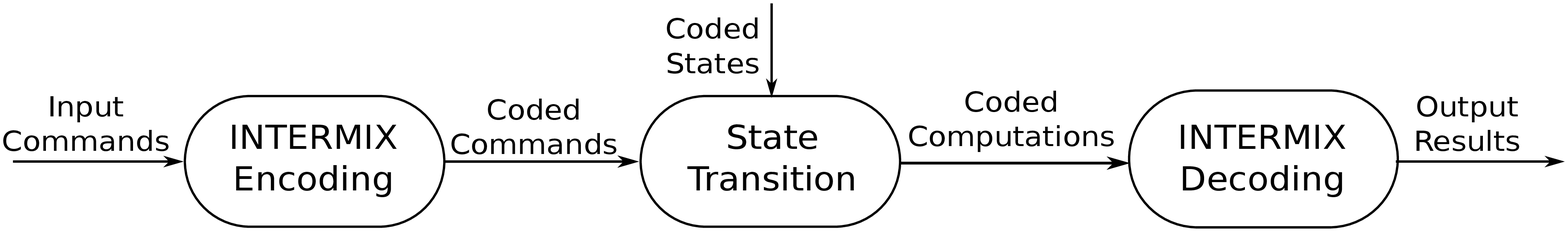}
  \caption{A block diagram illustration of main operations in Coded State Machine.}
  \vspace{-2mm}
  \label{fig:CSM}
\end{figure}


\bb{CSM works on a general class of state transition functions that are multivariate polynomials. The key idea behind CSM is to have each node execute the state transaction function $f$ on a {\em coded} state and a {\em coded} command, so that the $N$ coded outputs can be used to decode the $K$ original outputs.} More specifically, CSM generates each coded state/command by evaluating a Lagrange polynomial at some points, such that the output of $f$ can be viewed as evaluating another polynomial of higher degree at the same point. Given enough evaluations (some of them could be erroneous) of this new polynomial, we employ efficient noisy polynomial interpolation algorithms (e.g., Reed-Solomon decoding) to decode the $K$ original outputs, providing the security guarantees of CSM. \bb{Moreover, since a coded state has the same size as an original state, CSM retains the optimal storage efficiency. Finally, to reduce the encoding/decoding overhead for throughput scaling, we utilize INTERMIX to delegate the coding operations of all nodes to a single worker node, where fast polynomial interpolation/evaluation algorithms can be applied to minimize computational complexity. We illustrate the main operation flow of CSM in Figure~\ref{fig:CSM}.}


\noindent {\bf Related Works.} Conventional SMR research focuses on  designing optimal protocols for the consensus phase (\bb{where the nodes aim to agree on which input command submitted by the clients should be used}) that tolerate the maximum number of faulty processors 
~\cite{lamport1982byzantine,lamport1998part,oki1988viewstamped,dwork1988consensus,castro1999practical,castro2002practical}. CSM uses the same consensus protocols to decide on the input commands. 

Concepts from coding theory have been utilized earlier to provide fault tolerance to state machine execution. Building upon prior work \cite{garg2010implementing,balasubramanian2013fault},  fused state machine is proposed in~\cite{balasubramanian2014fault}. There, given $n$ primary finite state machines, $f$ fusion replica machines are constructed to correct up to $\frac{f}{2}$ Byzantine faults. The fusion replicas are components of the closed partition lattice~\cite{Hartmanis:1966:AST:1096925,lee2002closed} of the reachable product state of the primary machines, and are selected such that the minimum distance between any two product states is at least $f+1$.
Compared with the proposed CSM, while the fused state machine also achieves security scaling, the storage size at each replica increases with the number of primary machines $n$, as opposed to a constant storage size of a single state in CSM. 
When the state machine operations are write and read, i.e., the state machine system emulates a fault-tolerant shared memory, erasure codes have been exploited to minimize the storage costs given the coexistence of multiple versions of the data object, while ensuring that a reader can decode an atomically consistent version \cite{hendricks2007low,aguilera2005using,cachin2006optimal,dobre2013powerstore,cadambe2016information,wang2018multi}.  

The coding design of CSM is inspired by recent developments in \emph{coded computing}~\cite{li2016fundamental,LMA16_unify,lee2018speeding,dutta2016short,yu2017polynomial,TLDK-ICML,yu2018lagrangeNIPS}, which leverage tools from coding theory to inject computation redundancy in order to improve latency, security, and privacy of distributed computing  applications. In particular, as similarly done in Lagrange Coded Computing~\cite{yu2018lagrangeNIPS} to create coded data batches, CSM creates and processes coded states/commands using the Lagrange polynomial. However, in contrast to one-shot computation on static data in~\cite{yu2018lagrangeNIPS}, the requirement of dynamically updating the local coded state that is compatible with the upcoming state transition poses new challenges in the design of CSM.

Compared to the existing verifiable matrix-vector multiplication schemes   \cite{elkhiyaoui2016efficient,zhang2014efficient,fiore2012publicly,zhang2017new,sahraei2019interpol}, INTERMIX has two advantages. First, it is information-theoretically secure, i.e., it is secure even if the nodes have unlimited computation power. Second, INTERMIX enables almost every node in the network (with the exception of a constant number of auditors) to verify the correctness of the results in constant time. By contrast, in the existing schemes the verification complexity is at least linear in the size of the output. INTERMIX has the shortcoming that it is an interactive algorithm. However, the small number of required interactions (logarithmic in the number of nodes) allows us to trade a low communication overhead in exchange for the aforementioned desirable properties.

%% file: setting.tex
\vspace{-2mm}
\section{System Description and Problem Formulation}\label{sec:setting}
\vspace{-2mm}
We define a deterministic state machine (SM) as a collection of an input alphabet $\cal{X}$, an output alphabet $\cal{Y}$, a state space $\cal{S}$, and a deterministic state transition function $f: \cal{S} \times \cal{X} \rightarrow \cal{S} \times \cal{Y}$, for some vector spaces $\cal{X}$, $\cal{Y}$, and $\cal{S}$ over some field $\mathbb{F}$. The state of the state machine evolves in discrete rounds. At round $t$, given the current state $S(t)$ and the input command $X(t)$, the state machine computes a state transition function $(S(t+1),Y(t)) = f(S(t),X(t))$ to generate an output response $Y(t)$, and transitions into the next state $S(t+1)$.

We consider operating $K$ independent state machines with the same state transition function on $N \geq K$ unreliable compute nodes (e.g., hosting $K$ database services on $N$ commodity machines in a datacenter). There are a total of $M$ clients, who submit their queries/commands to the $K$ state machines for processing. In particular, at time $t$, each SM $k$ has state $S_k(t)$, and selects an input command $X_k(t)$ from the pool of commands submitted to SM $k$. Then, SM $k$ executes $X_k(t)$ by computing $(S_k(t+1),Y_k(t)) = f(S_k(t),X_k(t))$, returns the output result $Y_{k}(t)$ to the client who submits $X_k(t)$, and transitions into next state $S_k(t+1)$.

\begin{figure}[htbp]
  \centering
  \includegraphics[width=0.8\textwidth]{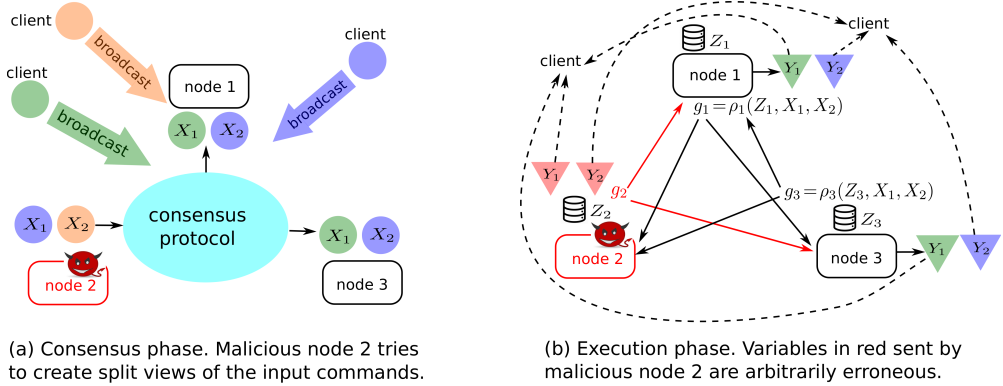}
    \vspace{-0.5mm}
  \caption{Illustration of operating $K\!=\!2$ state machines over $N\!=\!3$ nodes, one of which (node 2) is malicious. Malicious nodes can compromise the system security in the consensus phase, in the execution phase when honest nodes try to decode the computation results, and when delivering the decoded outputs to the clients.}
  \vspace{-1mm}
  \label{fig:setting}
\end{figure}

\vspace{-2mm}
\subsection{Network and Failure Models}
  \vspace{-2mm}
We consider a fully connected network between the clients and the compute nodes;  Figure~\ref{fig:setting}(a) shows  a subset ${\cal M}_k \subseteq \{1,\ldots,M\}$ of clients continuously and concurrently submit their commands to be processed at SM~$k$ by broadcasting them to all compute nodes. For the communication between the nodes, we study two settings:

    \noindent {\em   Synchronous network}, with a fixed and \emph{known} upper bound on the communication latency between any pair of nodes.
    
\noindent {\em  Partially synchronous network}, with unbounded communication delay until an \emph{unknown} global stabilization time (GST), after which the network becomes synchronous. Here a node expecting a message cannot distinguish between a failed message sender and a slow network since it does not know if  GST has been reached. 
 
To implement the state machine operations over the $N$ nodes, at round $t$, each node~$i$ locally stores some (possibly coded) data $Z_i(t) \in \mathbb{W}$, for some vector space $\mathbb{W}$ over $\mathbb{F}$, which is generated by some function $\phi_{i}^t$ over the states $S_1(t),\ldots,S_K(t)$. With these locally stored data, the implementation proceeds in two phases: the {\em consensus phase}, and the {\em execution phase}. 

\noindent {\bf Consensus Phase.} For each round index $t$, all network nodes run some consensus protocol, via exchanging messages (possibly over many iterations), to reach an agreement on a set of input commands $X_1(t),\ldots,X_K(t)$ to process (the consensus setting is standard in the literature,  e.g.,~\cite{dolev1987minimal}).  For each $k$, we label the index of the client who submits $X_k(t)$ as $m_k^t$.

\noindent {\bf Execution Phase.} As shown in Figure~\ref{fig:setting}(b), each node $i$ computes locally some intermediate result $g_i^t$, as some function $\rho_i^t$ of its stored data $Z_i(t)$, and the commands $X_1(t),\ldots,X_K(t)$ agreed in the consensus phase, and multicasts $g_i^t$ to a subset of other nodes. Using the local computation result, and the results received from other nodes, each node $i$ recovers the state transition functions for a subset ${\cal K}_i^t \subseteq \{1,\ldots,K\}$ of state machines via some function $\psi_{i}^t$, obtaining $|{\cal K}_i^t|$ estimates  $\psi_{i}^t(\{g_{i}^t\}_{i=1}^N) =\{(\hat{S}_{ik}(t+1),\hat{Y}_{ik}(t))\}_{k \in {\cal K}_i^t}$. Then, each node $i$ returns each of its computed output $\hat{Y}_{ik}(t)$ to the intended client $m_k^t$, and updates its local storage to $Z_i(t+1)$ by computing some function $\chi_{i}^t$ on the current storage 
$Z_i(t)$, and the updated states $\{\hat{S}_{ik}(t+1): k \in {\cal K}_i^t\}$. That is,
\begin{align}
    Z_i(t+1) = \chi_{i}^t(Z_i(t),\{\hat{S}_{ik}(t+1): \!k\! \in {\cal K}_i^t\}) = \phi_{i}^{t+1} (\{\hat{S}_{ik}(t+1)\!:\! k \in {\cal K}_i^t\}, \{\hat{S}_{ik}(t)\!:\! k \notin {\cal K}_i^t\}).
\end{align}
Finally, for each $k=1,\ldots,K$, after receiving the computed outputs $\{\hat{Y}_{ik}(t)\}_{i:k \in {\cal K}_i^t}$ from different nodes, client $m_k^t$ decides on the output $\hat{Y}_k(t)$. 

We consider an untrusted network where a subset of the nodes are subject to \emph{authenticated Byzantine faults}. That is, a faulty node can exhibit arbitrary behaviors that deviate from the above described protocol, but all messages between nodes are cryptographically signed, and hence impersonating others' messages is easily detectable. 

We say that a computation scheme $P$, specified by the nodes' storage design, and the node operations in the consensus and execution phases, is $b$-secure if for any subset ${\cal F} \subset \{1,\ldots,N\}$ with $|{\cal F}| \leq b$ such that the nodes in ${\cal H} = \{1,\ldots,N\} \setminus {\cal F}$ are honest, and the nodes in ${\cal F}$ are subject to authenticated Byzantine faults, the scheme achieves: 
    
    \noindent {\em Validity}: the command $X_k(t)$ selected in the consensus phase is indeed submitted by some client to SM $k$ before the start of round $t$, for all $k=1,\ldots,K$.
    
    \noindent {\em Consistency}: for each round
    index $t$, no two honest nodes in ${\cal C}$ decide on different values for $(X_1(t),\ldots,X_K(t))$.
   
   \noindent {\em Correctness}: for each $i \in {\cal C}$, $\hat{S}_{ik}(t+1) = S_k(t+1)$, for all $k \in {\cal K}_i$; and the client output $\hat{Y}_k(t) = Y_{k}(t)$, for all $k =1,\ldots,K$.
   
    \noindent {\em Liveness}: all clients' commands are executed.
  \vspace{-2mm}
\subsection{Performance Metrics}
  \vspace{-2mm}
A computation scheme $P$ is characterized by its security and efficiency. 

\noindent {\em Security} ($\beta_P$) is the maximum value of $b$ such that the computation scheme $P$ is $b$-secure.


\noindent {\em Storage efficiency}  
($\gamma_P$) is the ratio between the required memory size to store all $K$ states and the size of the data stored at each node, i.e., $    \gamma_P \triangleq \frac{K \log |\mathbb{S}|}{\log |\mathbb{W}|}$. 
Given a fixed storage size at each node, $\gamma_P$ indicates the maximum number of state machines the scheme $P$ can securely support.

\noindent {\em Throughput}  
($\lambda_P$)  is the average number of input commands that can be securely processed per unit operation at each node.
That is, $\lambda_P \triangleq  \liminf_{t \rightarrow \infty}\frac{K}{\sum_{i=1}^N (c(\rho_i^t) + c(\psi_i^t) + c(\chi_i^t))/N }$,
where $c(h)$ denotes the computational complexity of the function $h$ measured in number of additions and multiplications in $\mathbb{F}$. Here we focus on the regime where the computation latency is dominated by operations in the execution phase. Since the consensus phase of later rounds can be performed in parallel with the execution phase of current round, we do not consider the computational complexity of the consensus phase in the throughput definition. Also, we consider the case where all operations are carried out in memory, and no disk I/O is needed.


%% file: conventional.tex
\vspace{-2mm}
\section{State Machine Replication and Information-Theoretic Limits}\label{sec:SMR}
\vspace{-2mm}
In this section, we analyze state machine replication (SMR) and the information-theoretic limits.

\noindent {\bf Full vs. Partial Replication.} \bb{A classic SMR is full replication, where} the state of every state machine is replicated across all $N$ nodes. 
For each round $t$, the $N$ nodes altogether run some consensus algorithm to reach an agreement on the value of a vector of inputs $(X_1(t),\ldots,X_K(t))$. It is clear that the validity requirement is satisfied 
since each node knows all the commands submitted to all $K$ state machines.

\noindent {\em Synchronous network.} We use the Byzantine generals protocol~\cite{lamport1982byzantine} in the consensus phase, where a unique set of commands are proposed by a leader node and disseminated across the network. 
With the protection of digital signatures, the consistency requirement can be satisfied for an arbitrary number $b < N$ of malicious nodes. 
In the execution phase, each honest node 
executes the agreed commands, and sends the outputs to intended clients. For a state transition function $f$ with constant complexity, 
the full replication scheme achieves a constant throughput of $\lambda_{\textup{full}} = O(1)$. 
Each client waits for $b+1$ matching responses from the nodes before it accepts the output result. Hence, the total number of nodes $N$ needs to be at least $2b+1$. This promises a security of $\beta_{\textup{full}} = \lfloor\frac{N-1}{2}\rfloor = \Theta(N)$.

\noindent {\em Partially synchronous network.} We employ the PBFT protocol~\cite{castro1999practical,castro2002practical} in the consensus phase, 
which requires as least $N = 3b+1$ nodes. The execution phase is the same as in the synchronous setting. 
In partially synchronous networks, 
the security drops to $\beta_{\textup{full}} =  \lfloor\frac{N-1}{3}\rfloor = \Theta(N)$.  

Alternatively, we can partially replicate each state machine in a disjoint subset of $q = N/K$ nodes. This yields a storage efficiency of $\gamma_{\textup{partial}} = K$, and a throughput of $\lambda_{\textup{partial}} = \Theta(K)$. Partial replication has the same security as full replication on $q$ nodes. That is, $\beta_{\textup{partial}} = \lfloor\frac{q-1}{2}\rfloor = \Theta(N/K)$ in a synchronous network, and  $\beta_{\textup{partial}} = \lfloor\frac{q-1}{3}\rfloor = \Theta(N/K)$ in a partially synchronous network.


\noindent {\bf Information-Theoretic Upper Bounds.} Since the aggregated storage size of the entire network has to be at least the size of all $K$ state machines, the storage efficiency $\gamma \leq N$. To process $K$ input commands, the state transition function has to be executed at least $K$ times across the network, and thus the throughput $\lambda \leq \Theta(N)$. Finally, the maximum number of malicious nodes any computation scheme can tolerate cannot exceed half of the network size $N$. Thus, the security $\beta \leq \frac{N}{2}$. \bb{By letting $K$ scale linearly with $N$}, the upper bounds on all the three metrics scale linearly with $N$. \bb{However, both full replication scheme and partial replication scheme make tradeoffs between their scaling. Next, we present our main results that this tradeoff is {\em not} fundamental, and we can simultaneously achieve optimal scaling for security, storage, and throughput.}


%% file: results.tex
\vspace{-2mm}
\section{Main Results}
\vspace{-2mm}
We focus on a general class of state transition functions that are multivariate polynomials of maximum degree $d$. For instance, updating the balance of a bank account is a linear function of the current balance and the incoming deposit/withdrawal. Moreover, when we represent the variables in a state machine as bit streams, we can model any function as a polynomial using the following result~\cite{zou2011representing}: any Boolean function $\{0,1\}^n \rightarrow \{0,1\}$ can be represented by a polynomial of degree $\leq n$ with at most $2^{n-1}$ terms. See Appendix~\ref{sec:field} for an explicit construction of this polynomial.

\begin{theorem}[synchronous setting]\label{thm:synchronous}
Consider a system of $N$ compute nodes connected by synchronous networks, up to $\mu$ (for some constant $0 \leq \mu < \frac{1}{2}$) fraction of which are subject to authenticated Byzantine faults. Over this system, there exist computation schemes that support operating up to $\left\lfloor \frac{(1-2\mu)}{d}N + 1 -\frac{1}{d}\right\rfloor = \Theta(N)$ state machines with state transition function $f$ of degree $d$, and simultaneously achieve 
\begin{align}
  \textup{Storage efficiency} \quad  \gamma& \!=\!  \left\lfloor \frac{(1-2\mu)}{d}N + 1 \!-\!\frac{1}{d}\right\rfloor \!=\! \Theta(N),\\
  \textup{Security} \quad   \beta &\!=\! \mu N \!=\! \Theta(N).
\end{align}
Additionally, for broadcast networks (i.e., malicious nodes cannot send different messages to different nodes), we can also simultaneously achieve
\begin{align}
\textup{Throughput} \quad   \lambda &=\Theta\left(\frac{N}{\log^2 N \log \log N}\right).
\end{align}
\end{theorem}

\begin{theorem}[partially synchronous setting]\label{thm:partialSynchronous}
Consider a system of $N$ compute nodes connected by partially synchronous networks, up to $\nu$ (for some constant $0 \leq \nu < \frac{1}{3}$) fraction of which are subject to authenticated Byzantine faults. Over this system, there exist computation schemes that support operating up to $\left\lfloor \frac{(1-3\nu)}{d}N + 1 -\frac{1}{d}\right\rfloor = \Theta(N)$ state machines with state transition function $f$ of degree $d$, and simultaneously achieve
\begin{align}
  \textup{Storage efficiency} \quad  \gamma&\!=\!   \left\lfloor \frac{(1-3\nu)}{d}N + 1 \!-\!\frac{1}{d}\right\rfloor \!=\! \Theta(N),\\
  \textup{Security} \quad   \beta &\!=\! \nu N \!=\! \Theta(N).
\end{align}
\end{theorem}

We prove the storage and security scaling results in Theorem~\ref{thm:synchronous}, and Theorem~\ref{thm:partialSynchronous} in Section~\ref{sec:CSM}. In particular, we present a coded computation scheme named ``Coded State Machine'' (CSM), which simultaneously achieves storage and security scaling with the network size. In Section~\ref{sec:throughput}, we complete the proof of Theorem~\ref{thm:synchronous} by developing an information-theoretically verifiable matrix-vector multiplication algorithm (abbreviated as INTERMIX), which significantly slashes the computational complexity of CSM, and helps to achieve throughput scaling.


\begin{remark}
The proposed CSM simultaneously achieves the information-theoretically optimal storage efficiency and security to within constant multiplicative gaps, for both synchronous and partially synchronous settings. It also achieves optimal throughput to within a logarithmic factor for synchronous networks.
\end{remark}

\begin{remark}
In sharp contrast to SMR, each CSM node stores a coded state that is a unique linear combination of the $K$ states, whose size is the same as a single state variable. In the execution phase, each node generates a coded input command by linearly combining the $K$ incoming commands, and then computes the state transition function directly on the coded state and command. Using these coded computation results from all nodes, a subset of which may be erroneous, each node recovers the output results and the updated states via error-correcting decoding.
\end{remark}



\begin{remark}
For the special case of linear state transition function (i.e., $d=1$), codes designed for distributed storage (see, e.g.,~\cite{dimakis2011survey,rashmi2011optimal}) can also be used to achieve similar scaling as CSM. However, CSM is designed for a much more general class of arbitrary multivariate polynomials, which cannot be handled by state-of-the-art storage codes. 
\end{remark}

%% file: coded_state_machine.tex
\vspace{-2mm}
\section{Description of Coded State Machine}\label{sec:CSM}
\vspace{-2mm}

\bb{CSM simultaneously achieves optimal scaling of security, storage efficiency, and throughput. It has two key components, including {\em Coded State} and {\em Coded Execution}. We first briefly discuss these two components, and then describe CSM in detail.}

\begin{figure}[htbp]
  \centering
  \includegraphics[width=\textwidth]{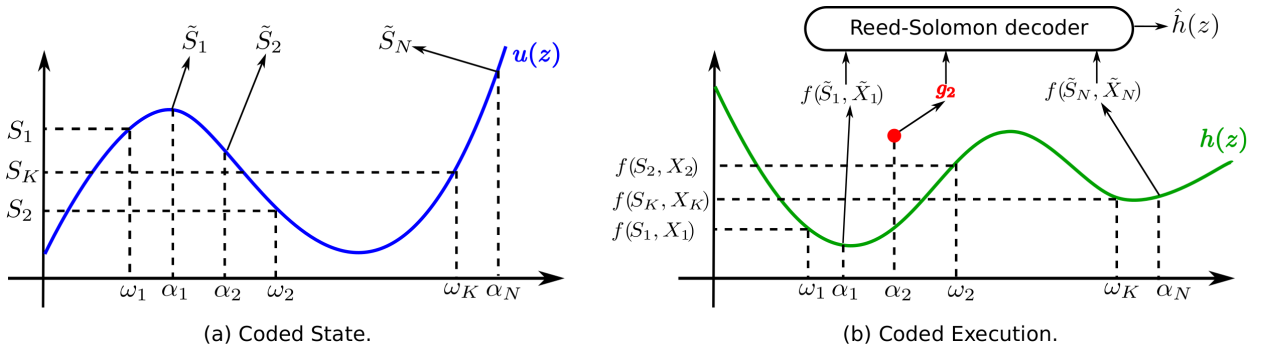}
  \caption{Illustration of the key components of Coded State Machine. (a) Coded State: each node $i$ stores a coded state $\tilde{S}_i$ generated by evaluating a polynomial $u(z)$ at point $\alpha_i$, where $u(\omega_k) = S_k$, for all $k=1,\ldots,K$. (b) Coded Execution: for a polynomial state transition function $f$, each honest node~$i$ computes an intermediate result $g_i = f(\tilde{S}_i,\tilde{X}_i)$, which can be viewed as evaluating a polynomial $h(z)$ with higher degree at point $\alpha_i$. Given $N$ computation results, a subset of which may be erroneous ($g_2$ in this case), Reed-Solomon decoding is used to recover $h$, which is then evaluated at $\omega_1,\ldots,\omega_K$ to obtain the final output results.}
  \label{fig:illustration}
\end{figure}

\noindent {\bf Coded State.} CSM views a (possibly coded) state variable as evaluating a polynomial $u(z)$ at a point. As illustrated in Figure~\ref{fig:illustration}(a), having specified $u(z)$ using the uncoded states $S_1,\ldots,S_K$ via Lagrange interpolation at distinct points $\omega_1,\ldots,\omega_K$, CSM generates $N$ coded states $\tilde{S}_1,\ldots,\tilde{S}_N$, by evaluating $u(z)$ at distinct points $\alpha_1,\ldots,\alpha_N$, and stores $\tilde{S}_i$ at node~$i$. This coding design was recently proposed in~\cite{yu2018lagrangeNIPS} for distributed computation of a polynomial over disjoint data.

\noindent {\bf Coded Execution.} In the execution phase, each node~$i$ generates a coded command~$\tilde{X}_i$ from the agreed commands $X_1,\ldots,X_K$ as done for the state encoding, and directly feeds $\tilde{S}_i$ and $\tilde{X}_i$ into the state transition function $f$. For any multivariate polynomial $f$, $f(\tilde{S}_i,\tilde{X}_i)$ can be viewed as evaluating a composite polynomial $h$ of higher degree at the point $\alpha_i$ (see Figure~\ref{fig:illustration}(b)). From the intermediate computation results $h(\alpha_1),\ldots,h(\alpha_N)$, a subset of which may be erroneous, CSM exploits efficient noisy interpolation techniques like Reed-Solomon decoding to recover $h$, and evaluates it at $\omega_k$ to obtain the output $f(S_k,X_k) = h(\omega_k)$, for all $k=1,\ldots,K$.

\subsection{Coded State}
\label{sec:codedstate}
To generate the coded states stored at the $N$ nodes, we first pick $K$ arbitrarily distinct elements $\omega_1,\ldots,\omega_K$ from the field $\mathbb{F}$, one for each state machine; and then pick $N$ arbitrarily distinct elements $\alpha_1,\ldots,\alpha_K$ from $\mathbb{F}$, one for each node.

Given the $K$ states $S_1(t),\ldots,S_K(t)$ at round $t$, we construct the Lagrange interpolation polynomial $u_t(z) = \sum_{k=1}^K S_k(t) \prod_{\ell \neq k} \frac{z - \omega_\ell}{\omega_k - \omega_\ell}$. Here evaluating $u_t(z)$ at $\omega_k$ recovers the state $S_k(t)$.

Then, 
the coded state $\tilde{S}_i(t)$ stored at node~$i$ is generated as evaluating $u_t(z)$ at the point $\alpha_i$, i.e., for all $i = 1,\ldots,N$,  
\begin{align}\label{eq:storeEncoding}
   \tilde{S}_i(t)= \sum_{k=1}^K S_k(t) \prod_{\ell \neq k} \frac{\alpha_i - \omega_\ell}{\omega_k - \omega_\ell} = \sum_{k=1}^K c_{ik}S_k(t),
\end{align}
where $c_{ik}$ is the coefficient for the state $S_k(t)$ at node $i$.

When $\mathbb{F}$ is finite, the field size $|\mathbb{F}|$ needs to be at least $N$ for this state encoding to be feasible. For small field (e.g., binary field), we can overcome this difficulty by using field extension and applying CSM on the extended field (see details in Appendix~\ref{sec:field}).

Since each coded state has the same size as an uncoded state, the CSM scheme has a storage efficiency of $\gamma_{\textup{CSM}} = K$.

\begin{remark}
The coefficients $c_{ik}$s in (\ref{eq:storeEncoding}) used to generate coded states are independent of both the state transition function $f$, and the round index $t$
Therefore, the state encoding of CSM is universally applicable for all types of state transition functions at each round of operations.
\end{remark}



\subsection{Coded Execution Phase}
\label{sec:codedexec}
The consensus phase is performed the same as SMR in Section~\ref{sec:SMR}, after which all honest nodes have agreed on the input commands $X_1(t),\ldots,X_K(t)$. As the first step in the execution phase, each node $i$ uses the same set of coefficients $c_{i1},\ldots,c_{iK}$ in (\ref{eq:storeEncoding}) to compute a coded command $\tilde{X}_i(t) = \sum_{k=1}^K c_{ik}X_k(t)$, which is the evaluation of a polynomial $v_t(z) = \sum_{k=1}^K X_k(t) \prod_{\ell \neq k} \frac{z - \omega_\ell}{\omega_k - \omega_\ell}$ at the point $\alpha_i$. Then, 
node $i$ applies the state transition function $f$ directly on $\tilde{X}_i(t)$ and its locally stored $\tilde{S}_i(t)$ to obtain $g_i^t = f(\tilde{S}_i(t),\tilde{X}_i(t))$, and broadcasts $g_i^t$ to all other nodes. Given that up to $b$ nodes are malicious, up to $b$ of the $N$ computation results are erroneous.

\begin{table}[htbp]
  \centering
\caption{Upper bounds on the number of malicious nodes ($b$) to achieve consensus on input commands, successful decoding, and secure delivery of output results.}
\label{table:bound}
  \begin{tabular}{| c | c | c | c |c| }
    \hline
    \rule{0pt}{10pt} & Input Consensus & Decoding & Output Delivery \\ \hline
    \rule{0pt}{10pt} Synchronous & $b+1 \leq N$  & $2b+1 \leq N-d(K-1)$  &  $2b+1 \leq N$  \\ \hline
    \rule{0pt}{10pt}  Partially Synchronous & $3b+1 \leq N$ & $3b+1 \leq N-d(K-1)$  & $2b+1 \leq N$  \\ \hline
  \end{tabular}
\end{table}

\noindent {\em Synchronous network.} 
Each node waits for a fixed interval to receive all the computation results $g_1^t,\ldots,g_N^t$. 
For a function $f$ that is a multivariate polynomial of constant degree $d$, the result $g_i^t$ can be viewed as the evaluation of a univariate polynomial $f(u_t(z),v_t(z))$ of degree $d(K-1)$ at $z=\alpha_i$. Given $N$ such evaluation results from distinct $\alpha_i$s, and $b$ of them might be arbitrarily erroneous, each node can recover the coefficients of the polynomial $f(u_t(z),v_t(z))$ following the procedure of decoding a Reed-Solomon code with dimension $d(K-1)+1$ and length $N$ (see, e.g.,~\cite{roth2006introduction}). This decoding can be successful if and only if the \# of errors is bounded as $2b \leq N - d(K-1)-1$. After decoding, each node $i$ locally reconstructs a polynomial $\hat{f}_i(u_t(z),v_t(z))$, and evaluates it at $\omega_k$ to obtain $\hat{f}_i(u_t(\omega_k),v_t(\omega_k)) = \hat{f}_i(S_k(t),X_k(t)) = (\hat{S}_{ik}(t+1),\hat{Y}_{ik}(t))$, for all $k=1,\ldots,K$. We note that the reconstructed polynomials at all honest nodes are identical even when the malicious nodes deliberately send different computation results to different nodes (i.e., in presence of equivocation). 

Finally, each node $i$ sends the recovered output $\hat{Y}_{ik}(t)$ to the intended client $m_k^t$, and updates its local storage to $\tilde{S}_i(t+1) = \sum_{k=1}^K c_{ik} \hat{S}_{ik}(t+1)$. 
In Table~\ref{table:bound}, we summarize the upper bounds on $b$ for the system to achieve consensus on input commands, successful decoding, and secure delivery of output results for synchronous networks, among which the one for successful decoding is most effective. Assuming a $\mu$ fraction of the $N$ nodes are malicious, i.e., $b = \mu N$, then this computing system can securely support up to $K \leq \frac{(1-2\mu)N-1}{d}+1$ state machines. 

\noindent {\em Partially synchronous network.} In this case, since each of the $b$ malicious nodes may refrain from sending any messages, the remaining honest nodes should start decoding upon receiving $N-b$ computation results to ensure liveness. However, since a receiver node cannot distinguish between a missing message held by a faulty node and a delayed message sent by an honest node, 
a node can proceed to the decoding step with up to $b$ of the received results being erroneous. In this case, each node needs to decode a Reed-Solomon code with dimension $d(K-1)+1$ and length $N-b$, and successful decoding requires $2b \leq N-b-d(K-1)-1$. 
Assuming a $\nu$ fraction of the $N$ nodes are malicious, i.e., $b = \nu N$, then this system can securely support up to $K \leq \frac{(1-3\nu)N-1}{d}+1$ state machines. This completes the proof of  Theorem~\ref{thm:partialSynchronous}.

%% file: throughput.tex
\section{Throughput Scaling via Verifiable Computing}\label{sec:throughput}
\begin{figure}
    \centering
    \includegraphics[scale=0.14]{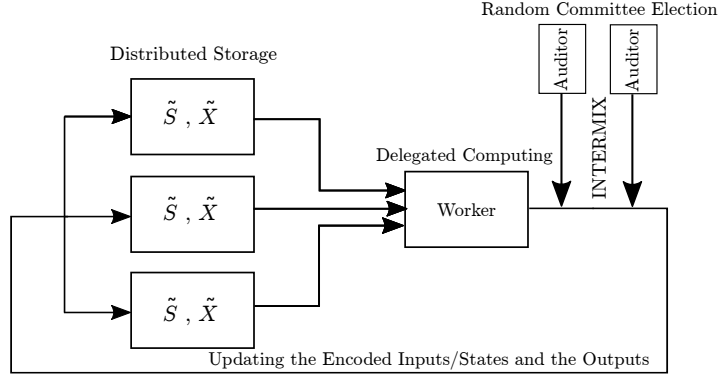}
    \caption{A block-diagram of the proposed  centralized computation model for throughput scaling. INTERMIX is used for verifying the correctness of the results.}
    \label{fig:block_diagram}
\end{figure}

As defined in Section~\ref{sec:setting}, the throughput of CSM is inversely proportional to $\sum_{i = 1 }^N c(\rho_i^t) + c(\psi_i^t) + c(\chi_i^t)$ where $c(\rho)$, $c(\psi)$, and $ c(\chi)$ respectively represent the computational complexities of encoding and processing the input commands, decoding the state transition functions, and updating the coded states. In this section we describe a protocol that allows for throughput scaling by reducing this overall complexity to quasilinear in $N$. Our protocol comprises of two main components. First, we describe INTERMIX, an efficient tool for information-theoretically verifiable matrix-vector multiplication, which is at the core of all our encoding and decoding operations. INTERMIX consists of a single worker that performs the computation, and a small randomly elected committee of auditors that are in charge of examining the worker's output. Once an honest auditor detects a fraud, he interactively enforces the worker to produce an inconsistent result that can be checked in constant time by every node in the network. Next, we show that by delegating the entire encoding and decoding operations to a single node in the network, the overall complexity can be reduced to quasilinear in $N$. Computing state transition on the coded state and commands will be done distributedly at each node, similar as before. We will show that the correctness of all the operations performed by this single worker can be checked by every node in the network, by using INTERMIX as a blackbox verification module. A block-diagram of the proposed model has been illustrated in Figure \ref{fig:block_diagram}. We note that the correctness of the results in this section relies on additional assumptions of synchronous and broadcast (no equivocation) network.

\subsection{INTERMIX, Efficiently Verifiable Matrix-Vector Multiplication}
Suppose we have a network of $N$ nodes, a matrix $A\in \mathbb{F}^{N\times K}$ and a vector $X\in \mathbb{F}^{K\times 1}$. Node $i$ is interested in computing $A_iX$ where $A_i$ represents the $i$'th row of $A$. Our objective is to delegate the entire computation to only one node (the worker), while the remaining nodes verify the correctness of the results. 
Our approach is to randomly choose $J$ nodes to audit the worker, where $J$ is a constant large enough that the probability that none of the auditors is honest is negligible. Each auditor will repeat the computation of $Y = AX$ and compares it with the result returned by the worker. An (honest) auditor who detects a fraud in the computation, will interactively enforce the worker to produce an inconsistent result which can be checked in constant time by the remaining nodes in the network. The algorithm has the following components.

\begin{figure}
    \centering
    \includegraphics[scale=0.4]{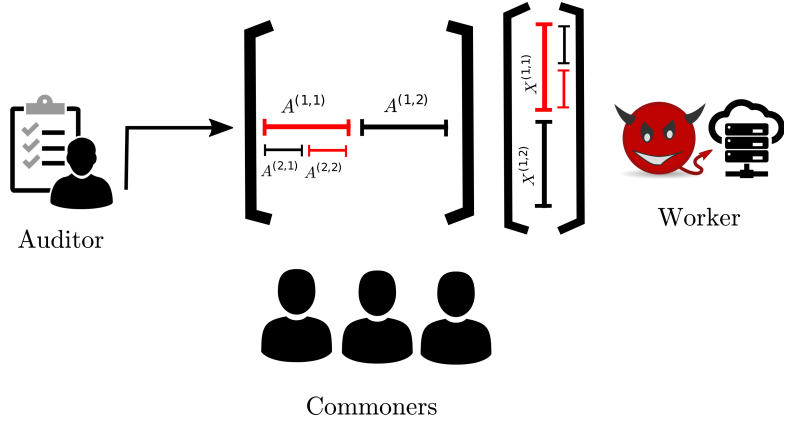}
    \caption{Illustration of INTERMIX for verifiable matrix vector multiplication. An honest auditor can interactively enforce the worker to provide an inconsistent result that can be checked in constant time. We assume that the remaining nodes (the commoners) can overhear the entire conversation.}
    \label{fig:INTERMIX}
\end{figure}

\begin{itemize}[leftmargin=*]
\item{\bf Random Committee/Leader Election}:
A worker and a committee of auditors are chosen randomly. Given that at most a $\mu$ fraction of the nodes in the network are dishonest, we choose $J = \log \epsilon /\log \mu$ auditors at random, in order to ensure with probability $1-\epsilon$, that at least one auditor is honest. An easy way to do this is to allow each node to self-elect with probability $\frac{J}{N}$. If an adversary decides to conduct a DoS attack by imposing unnecessary audits, he will be banned from the next rounds of the algorithm. 

In a dynamic setting, where banning is ineffective, the local coin-toss algorithm can be replaced by any off-the-shelf distributed random number  generating algorithm (such as \cite{syta2017scalable,popov2017decentralized}). Given that we do not need {\it fresh} randomness at every round of computation, we only run such an algorithm occasionally to amortize its overhead. The committee can still be updated at every round by relying on a standard pseudorandom number generator.

We can further hinder a dynamic adversary who wishes to corrupt the auditors {\it after} the election process, by keeping the identities of the auditors secret with the help of Verifiable Random Functions (VRF) \cite{dodis2005verifiable,micali1999verifiable}. Accordingly, an auditor remains anonymous until he decides to conduct an audit, at which point he presents a proof of having been elected.

\begin{remark} It is important to note that the auditors are stateless, which makes it possible to change the committee at every round of computation, without imposing significant communication overhead on the network. To make a comparison, random sortition algorithms \cite{kokoris2017omniledger,naor2003simple} are more prone to dynamic adversaries, due to the prohibitively large communication overhead  associated with frequently updating the allocation of the nodes to the state machines.
\end{remark}

\item {\bf Computation (Auditors and the Worker)}: If the worker is honest, he computes $\hat{Y} = AX$ and broadcasts this result to the network. If he is dishonest, he broadcasts an arbitrary $\hat{Y}$. Each auditor repeats the computation of $AX$. 

\item {\bf Interactive Localization of the Fraud (Auditors and the Worker)}: If an (honest) auditor does not detect a fraud in the computation, he will inform the rest of the network (henceforth referred  to as the commoners) that the result is correct. Otherwise, he will aim at proving to the commoners that the result returned by the worker is wrong. He will accomplish this task through a set of $\log K$ interactive queries to the worker as follows.

Note that if $\hat{Y}\neq Y$, there must exist at least one index $i\in \{1,\ldots,N\}$ such that $Y_i\neq \hat{Y}_i$. If there are multiple such indices, the auditor will choose one such $i$ arbitrarily, and will aim at proving to the commoners that $\hat{Y}_i\neq A_iX$, which will be sufficient to prove that $\hat{Y} \neq Y$. To fulfill this goal, first the auditor breaks $A_i$ into two consecutive chunks of equal size named $B_1$ and $B_2$. Similarly, the vector $X$ is broken into $X_1$ and $X_2$. The auditor asks the worker to compute $Z_1 = B_1X_1$ as well as $Z_2 = B_2X_2$. If $\hat{Z}_1+ \hat{Z}_2 \neq \hat{Y}_i$, the auditor raises an alert to draw the attention of the commoners to this fact. A commoner can simply check that $\hat{Z}_1+ \hat{Z}_2 \neq \hat{Y}_i$ and is convinced of the fraud. On the other hand, if $\hat{Z}_1+ \hat{Z}_2 = \hat{Y}_i$, we must have either $\hat{Z}_1\neq B_1X_1$ or $\hat{Z}_2\neq B_2X_2$. The auditor proceeds to compute both $ Z_1$ and $Z_2$ in order to locate the fraud. Once this is done, the auditor can focus on the corresponding half of the vector and repeat the algorithm, until an inconsistency is detected. This algorithm is summarized below.
\begin{algorithm}
\caption[caption]{Algorithm run by an honest auditor}
\begin{algorithmic}[1]
\Statex {{\bf Input:} $A, X, \hat{Y}$.}
 \Statex {\bf Output: } One bit indicating whether the auditing process has succeeded or failed. A string $\zeta$ that localizes the failure.
\If{$\hat{Y} = AX$}
\State Return True.
\EndIf
\State Choose $i$ such that $\hat{Y}_i \neq A_iX$. Set $\zeta = [i]$.
\State  Set $j = 1$, $A^{(1)} = A_i$, $X^{(1)} = X$,  $\hat{Y}^{(1)} = \hat{Y}_i$, $s = \mbox{length}(A^{(1)})$.
\While{($s > 1$)}
 \State Let $A^{(j,1)} = A^{(j)}_{1:\lfloor s/2\rfloor}$, $A^{(j,2)} = A^{(j)}_{\lfloor s/2\rfloor + 1:s}$, $X^{(j,1)} = X^{(j)}_{1:\lfloor s/2\rfloor}$, \hphantom{1cm}$X^{(j,2)} = X^{(j)}_{\lfloor s/2\rfloor + 1:s}$. 
 \State Request $Y^{(j,1)} = A^{(j,1)}X^{(j,1)}$ and $Y^{(j,2)} =A^{(j,2)}X^{(j,2)}$ from \phantom{1cm}the worker. Receive $\hat{Y}^{(j,1)}, \hat{Y}^{(j,2)}$.
\If{ $\hat{Y}^{(j,1)}+\hat{Y}^{(j,2)}\neq\hat{Y}^{(j)}$}
\State Return $($False$,\zeta)$.
\Else
\State Choose $\ell\in\{1,2\}$ such that {$\hat{Y}^{(j,\ell)} \neq {Y}^{(j,\ell)}$}. Let $s = \hphantom{1.5cm}$\;\;$  \mbox{length}(Y^{(j,\ell)})$ and $\zeta = [\zeta, \ell]$.
\State Let $\hat{Y}^{(j+1)} = \hat{Y}^{(j,\ell)}$, $A^{(j+1)} = A^{(j,\ell)}$ and $X^{(j+1)} = X^{(j,\ell)}$.
\State Increment $j$.
\EndIf
\EndWhile
\State Return $($False$,\zeta)$.
\end{algorithmic}
\label{Alg:interactive_localization}
\end{algorithm}

\item {\bf Verification (Commoners)}: If all the auditors acknowledge the correctness of the result, the commoners will accept the result provided by the worker. If an auditor returns False in Algorithm \ref{Alg:interactive_localization} within the loop (while $s>1$), the commoners will check in constant time whether $\hat{Y}^{(j,1)} + \hat{Y}^{(j,2)}= \hat{Y}^{(j)}$. If this equality does not hold, the commoners will conclude that the result is incorrect. Finally, if the algorithm returns False in the last stage, the commoners will check in constant time whether $\hat{Y}^{(j)}= A^{(j)}X^{(j)}$ holds or not.
\end{itemize}

\noindent {\bf Correctness.} If the worker is honest, an honest auditor will observe that $\hat{Y} = Y$ and acknowledge its correctness. If all the auditors return True, the commoners will accept the result as correct. The action space of a dishonest auditor is quite limited. Firstly, he can impose unnecessary audits on the honest worker despite observing that $\hat{Y} = Y$. Furthermore, he can return False despite detecting no inconsistency in the responses to the audits. By doing so, he can merely worsen the complexity of the system. A commoner can verify that there is no inconsistency in the outputs of the worker in constant time and dismiss the malicious auditor's alert.   

\noindent {\bf (Information-theoretic) Soundness.} Suppose the worker is dishonest, i.e., $\hat{Y} \neq Y$. Remember that we chose $J$ large enough that the probability of having no honest auditor is $\epsilon$. Therefore, with probability $1-\epsilon$ there will be an honest auditor which will be able to localize the fraud of the worker following the interactive method described in Algorithm \ref{Alg:interactive_localization}. Furthermore, relying on the assumptions of broadcast and synchronous network, if a worker chooses to not respond to any auditor, the commoners can detect this violation of the protocol, and decide that the worker is malicious. Note that the soundness of INTERMIX does not rely on any assumption on the computation power of the worker. In other words, INTERMIX is information-theoretically secure, since even a computationally unbounded worker is unable to compromise the soundness of the algorithm.

\noindent {\bf Overall Computational Complexity of the Verification Scheme.} Let us compute the overall complexity under a worst case assumption that all the auditors conduct queries to the worker. One auditor can increase the complexity of the worker and himself by $2r_{K/2} + 2r_{K/4} + 2r_{K/8}+\cdots$ where $r_j$ represents the complexity of computing the inner-product of two vectors of length $j$, and is equal to $2j$. As a result, in a worst case scenario, where all the auditors conduct queries, the overall added complexity due to inner-product computation is $8JK$. Each auditor also performs $3\log(K)$ comparisons between his locally computed results and the results returned by the worker.

A auditor can also increase the complexity of the commoners by returning False in Algorithm \ref{Alg:interactive_localization}. In this case, a commoner will conduct one comparison operation to check if the auditor's alert is valid or not. In a worst case scenario, where all the auditors return False, the overall complexity of the commoners increases by $N-J-1$.

Therefore, the worst-case computational complexity of INTERMIX is $(J+1)c(AX) + 8JK +3J\log K + N- J -1 $, where $c(AX)$ denotes the computational complexity of multiplying $A$ by $X$. We observe that as $N$ grows large, and unless the matrix $A$ has a very simple structure, the overall complexity of INTERMIX is dominated by the complexity of centralized  computation of $AX$.
\subsection{Centralization of Encoding/Decoding for Throughput Scaling}
To reduce the coding complexity using the idea of verifiable computing, we must address a fundamental question: {\it if all encoding/decoding operations are performed at a single node, can we achieve a per node complexity that scales sub-linearly with $N$?} To answer this question, let us look at all the coding operations in CSM, i.e., {\it (i) encoding of input commands,
     (ii) decoding of output results/next states, and
 (iii) updating coded states.} We present computation schemes at a single node that has sub-linear complexity in $N$ for each of the above coding operations. We will use INTERMIX as a verification module throughout this section, wherever the need for trusted computation arises.
 
\noindent {\bf Encoding of input commands.}
As we saw in Section \ref{sec:codedexec}, for the encoding of the input commands, we pick $K$ distinct elements $\omega_1,\ldots,\omega_K \in \mathbb{F}$, and $N$ distinct elements $\alpha_1,\ldots,\alpha_N \in \mathbb{F}$. At each round $t$, we perform a Lagrange polynomial interpolation using the points $(\omega_1,X_1(t)),\ldots,(\omega_K,X_K(t))$ to construct a polynomial $v_t(z)$. 
For each node~$i$, the coded command $\tilde{X}_i(t)$ is generated as $\tilde{X}_i(t) = v_t(\alpha_i)$.
We consider computing coded commands of all $N$ nodes at a single node which we call the worker. The encoding process proceeds in two steps: 1) polynomial interpolation, 2) multi-point polynomial evaluation.

\noindent {\bf Step 1: Polynomial interpolation.} Given the input commands $X_1(t),\ldots,X_K(t)$, the worker performs Lagrange interpolation to find a polynomial $v_t(z)$ of degree $K-1$ that passes through all $X_i(t)$, i.e., $    v_t(z) = \sum_{k=1}^K X_k(t) \prod_{\ell \neq k} \frac{z - \omega_\ell}{\omega_k - \omega_\ell} = \sum_{j=0}^{K-1} a_j(t)z^j$,
where $a_j(t)$ is the coefficient for the term $z^j$. This operation can be done with computational complexity (operation counts) of $O(K \log^2 K \log \log K)$ (see, e.g.,~\cite{kedlaya2011fast}).

\noindent {\bf Step 2: Multi-point polynomial evaluation.} Having obtained the coefficients $a_0,\ldots,a_{K-1}$, the worker evaluates $v_t(z)$ at the points $\alpha_1,\ldots,\alpha_N$ to compute the coded inputs $\tilde{X}_1(t),\ldots,\tilde{X}_N(t)$. Specifically, $\begin{bmatrix}\tilde{X}_1(t) \\ \vdots \\ \tilde{X}_N(t) \end{bmatrix}= \begin{bmatrix}1 & \alpha_1 & \cdots & \alpha_1^{K-1} \\ \vdots & \vdots & \ddots & \vdots \\
1 & \alpha_N & \cdots & \alpha_N^{K-1}\end{bmatrix} \begin{bmatrix}a_0(t) \\ \vdots \\ a_{K-1}(t) \end{bmatrix} =  \begin{bmatrix}c_{11} & \cdots & c_{1K} \\ \vdots & \ddots & \vdots \\
c_{N1} & \cdots & c_{NK}\end{bmatrix} \begin{bmatrix}X_1(t) \\ \vdots \\ X_{K}(t) \end{bmatrix}$,
where $c_{ik} = \prod_{\ell \neq k} \frac{\alpha_i - \omega_\ell}{\omega_k - \omega_\ell}$ is the coefficient of $X_k(t)$.  This operation can be done using fast polynomial arithmetic with computational complexity $O(N \log^2 N \log \log N)$ (see, e.g., ~\cite{kedlaya2011fast,macwilliams1977theory}). Hence, the total computational complexity to encode the input commands is $O(N \log^2 N \log \log N)$. Define $C = [c_{ik}]_{N\times K}$. The committee of auditors will use the second equality, i.e., fact that $\tilde{X}(t) = CX(t)$ to interactively verify the correctness of the results with the INTERMIX algorithm.

\noindent {\bf Updating coded states.} At the end of round~$t$, each node~$i$ updates its local coded state to $\tilde{S}_i(t+1)$. Centralized state update can be done similarly to encoding the input commands. Here, the worker interpolates the polynomial $u_{t+1}(z)$, as defined in Section \ref{sec:codedstate}, and evaluates it at $\alpha_1,\dots,\alpha_N$. 
The auditors take advantage of the fact that $\begin{bmatrix}\tilde{S}_1(t+1) \\ \vdots \\ \tilde{S}_N(t+1) \end{bmatrix} =  \begin{bmatrix}c_{11} & \cdots & c_{1K} \\ \vdots & \ddots & \vdots \\
c_{N1} & \cdots & c_{NK}\end{bmatrix} \begin{bmatrix}S_1(t+1) \\ \vdots \\ S_{K}(t+1) \end{bmatrix}$
to help the commoners verify the correctness of the results in constant time, via INTERMIX.
\noindent {\bf Decoding of the output results/new states.} Consider a state transition function $f$ of degree $d$. The coded computation result at an honest node~$i$ can be viewed as evaluating a polynomial $h_t$ of degree $K'=(K-1)d$ at the point $\alpha_i$. That is, $   f(\tilde{X}_i(t),\tilde{S}_i(t)) = h_t(\alpha_i) = \sum_{j=0}^{K'} b_j(t) \alpha_i^j$,
where $b_j(t)$ is the coefficient of the term $z^j$ in $h_t$. After receiving the computation results $\{g_i^t\}_{i=1}^N$ from all $N$ nodes, up to $\mu$ fraction of which are erroneous, a worker node decodes $b_0(t),\ldots,b_{K'}(t)$ with a computational complexity of $O(N \log^2 N \log \log N)$ (say, using Berlekamp-Welch algorithm). 


Having decoded $b_0(t),\ldots,b_{K'}(t)$, the worker node is required to broadcast these coefficients to the rest of the network, and then evaluates $g$ at the points $\omega_1,\ldots,\omega_K$ to recover the output results and the next set of states. That is, the worker node computes
\begin{align}
    \begin{bmatrix}(Y_1(t),S_1(t+1)) \\ \vdots \\ (Y_K(t),S_K(t+1)) \end{bmatrix}= \begin{bmatrix}1 & \omega_1 & \cdots & \omega_1^{K'} \\ \vdots & \vdots & \ddots & \vdots \\
1 & \omega_K & \cdots & \omega_K^{K'}\end{bmatrix} \begin{bmatrix}b_0(t) \\ \vdots \\ b_{K'}(t) \end{bmatrix}.\label{eq:centralized_transition_output}
\end{align}
The computational complexity of this step is $O(N \log^2 N \log \log N)$. 

Two steps are required to verify the correctness of the decoded results. Firstly, the worker needs to prove that his decoding of the coefficients $b_0(t),\dots,b_{K'}(t)$ are correct. Secondly, he needs to prove that \eqref{eq:centralized_transition_output} is computed correctly. This second step can be directly accomplished via INTERMIX applied on the matrix $\Omega=[\omega_i^j]_{K\times (K'+1)}$ and the vector $[b_0(t) \dots b_{K'}(t)]^\top$. We will now describe how INTERMIX can be used for verifiable polynomial interpolation in the presence of errors.  

We know from principles of error correction coding, that the polynomial $h_t(z) = \sum_{j = 0}^{K'} b_j(t)z^j$ is the correct interpolation among the $N$ received points $\{(\alpha_i,g_i^t)\}_{i=1}^N$, if and only if there exists a set $\tau$ of size at least $N - \frac{N - K' - 1}{2} = \frac{N+K'+1}{2}$ such that $h_t(\alpha_i) = g_i^t$ for all $i\in \tau$. We will require the worker to broadcast this set $\tau$ together with the decoded coefficients $b_0(t),\dots,b_{K'}(t)$. Let us without loss of generality assume that $\tau = \{1,\ldots,\frac{N + K' + 1}{2}\}$. It must hold that 
\begin{eqnarray}
\begin{bmatrix}
g_1^t \\ \vdots \\ g_{\frac{N+K'+1}{2}}^t
\end{bmatrix}
= \begin{bmatrix}1& \alpha_1 &\dots & \alpha_1 ^{K'} \\ \vdots &\vdots & \ddots & \vdots\\ 1 & \alpha_{\frac{N+K'+1}{2}}& \dots & \alpha_{\frac{N+K'+1}{2}}^{K'} \end{bmatrix}
\begin{bmatrix}
b_0(t) \\ \vdots \\ b_{K'}(t)
\end{bmatrix}.
\label{eq:interpolation_trick}
\end{eqnarray}
Now, we can apply INTERMIX to verify the correctness of \eqref{eq:interpolation_trick}, and subsequently verify the correctness of the decoding operation.

\subsection{Evaluating the Throughput} 
We are now ready to characterize the throughput of CSM using INTERMIX, and establish the final claim of Theorem \ref{thm:synchronous}. We saw in this section that the complexities of the 
encoding and decoding operations are reduced to $O(N\log^2 N \log\log N)$ given that the entire computation is delegated to only one node. Furthermore, each intermediate result can be computed locally in constant time given that the polynomial $f(\tilde{X}_i(t),\tilde{S}_i(t))$ has a fixed degree that does not grow with $N$. 
Therefore, the overall computational complexity of the CSM is $\sum_{i=1}^N c(\rho_i^t) + c(\psi_i^t)+ c(\chi_i^t) = O(N\log^2N\log\log N) + O(N)$,
where the first term indicates the complexity of the auditors and the worker, and the second term indicates the aggregate complexity of the remaining nodes. The throughput of CSM is computed as $    \lambda_{\textup{CSM}} =  \liminf_{t \rightarrow \infty}\frac{K}{\sum_{i=1}^N (c(\rho_i^t) + c(\psi_i^t) + c(\chi_i^t))/N } = \Theta\left(\frac{N}{\log^2 N \log\log N}\right)$. 
This completes the proof of the last statement about throughput scaling in Theorem \ref{thm:synchronous}.

%% file: discussion.tex
\vspace{-2mm}
\section{Discussions}
\vspace{-2mm}
In this section, we discuss engineering considerations and future research directions raised by CSM. 

\noindent {\bf Blockchain Applications.} One motivation for working on CSM is that (sharded) blockchain systems are best represented using the state machine formalism. While the current instantiations of blockchains are based on full replication, future proposals are most similar to partial replication \cite{luu2016secure}, both of which make severe security-efficiency tradeoffs~\cite{blogpost}. The results of this paper on coded state machines can be used as a stepping stone towards scalable and secure blockchain designs. 

\noindent {\bf Random Allocation vs. CSM.} An alternate architecture for scaling security and efficiency simultaneously is to randomly allocate nodes into groups that process distinct state machines. In this method, the fraction of adversaries in the group will be representative of the fraction of adversaries in the entire network. However, a dynamic adversary who observes this grouping can do post-facto corruptions to the small number of nodes in that group, thus making security proportional to group size. One possible solution is to rotate the group allocations periodically \cite{kokoris2017omniledger}, but this cannot be very frequent since this will require each node to re-download the data corresponding to different state machines. In CSM we avoid this tradeoff completely, and the full security is guaranteed against a dynamic adversary as well.

\noindent {\bf Verifiable Computing vs. INTERMIX.} In order to scale throughput, there have been several existing proposals to do verifiable computing \cite{gennaro2010non,bitansky2012extractable,parno2016pinocchio,ben2014succinct,ben2018scalable}, where one node does computation and other nodes verify the computation in sub-linear time. These methods are only secure under cryptographic assumptions, are not yet practically scalable and potentially require a trusted setup.  By contrast, INTERMIX is information-theoretically secure, is practical and does not require a trusted setup. However, compared with some other verifiable computation approaches, the interactive nature of INTERMIX becomes a disadvantage.

\noindent {\bf Degree Dependence.} The proposed CSM is efficient only when the degree of the polynomial is a constant. As a future direction of research, we aim at generalizing the results to the scenario where the state machine can be represented as a low-depth arithmetic circuit. While low-depth circuits can still have high polynomial degrees, their algebraic structures can be potentially exploited to design efficient computation schemes.

\noindent {\bf Distinct State Machines.} Our present approach in CSM assumes that the $K$ distinct state machines have the same state transition function, only distinct sequences of inputs. A future direction of research will be to generalize CSM to the case of distinct state transition functions.

%% file: app.tex
\section{Field Extension for General Boolean Functions}\label{sec:field}

Using the construction of~\cite[Theorem 2]{zou2011representing}, we can represent any arbitrary Boolean function $f: \{0,1\}^{n} \rightarrow \{0,1\}$ whose inputs are $n$ binary variables as a multivariate polynomial $p$ of degree $n$ as follows. For each vector $\mathbf{a} = (a_1,\ldots,a_n) \in \{0,1\}^n$, we define $h_{\mathbf{a}} = z_1z_2\cdots z_n$, where $z_i = x_i$ if $a_i=1$, and $z_i=y_i$ if $a_i=0$. Next, we partition $\{0,1\}^n$ into two disjoint subsets $S_0$ and $S_1$ as follows.
\begin{align}
    S_0 &= \{\mathbf{a} \in \{0,1\}^n: f(\mathbf{a}) =0\},\\
    S_1 &= \{\mathbf{a} \in \{0,1\}^n: f(\mathbf{a}) =1\}.
\end{align}
The polynomial $p$ is then constructed as 
\begin{align}
    p(x_1,\ldots,x_n, y_1,\ldots,y_n) \!=\! \sum_{\mathbf{a} \in S_1} h_{\mathbf{a}} \!=\! 1 \!+\! \sum_{\mathbf{a} \in S_0} h_{\mathbf{a}},\label{eq:boolPoly}
\end{align}
where $y_i = x_i+1$.


For a state machine whose state, input command, and output result are represented as bit streams, with a state transition polynomial $p$ over binary field (i.e., $\mathbb{F} = \{0,1\}$) in the form of~(\ref{eq:boolPoly}), the state encoding in~(\ref{eq:storeEncoding}) does not directly apply since it requires the field size $|\mathbb{F}|$ to be at least the network size $N$. To use CSM in this case, we can embed each element $s_{k}[i] \in \{0,1\}$ of a state $S_k$ (time index omitted) into a binary extension field $\mathbb{F}_{2^m}$ with $2^m \geq N$. Specifically, the embedding $\bar{s}_k[i] \in \mathbb{F}_{2^m}$ of the element $s_k[i]$ is generated such that
\begin{align}\label{eq:extension}
\bar{s}_k[i]= \begin{cases} \underbrace{00\cdots0}_{m}, & s_k[i]=0,\\
\underbrace{00\cdots0}_{m-1}1, & s_k[i]=1.
\end{cases}
\end{align}
Then we can select distinct elements $\alpha_1, \ldots, \alpha_N \in \mathbb{F}_{2^m}$ to apply the encoding on the states in the extension field. We also use the same embedding and coding on the input commands.

State transition over extension field generates the correct result. To see that, we can easily verify that the value of the state transition polynomial $p$ as constructed in (\ref{eq:boolPoly}) is invariant with the embedding operation in (\ref{eq:extension}). That is, since the polynomial $p$ is the summation of monomials in $\mathbb{F}_2$, when we replace each input bit with its embedding, the output value is $\underbrace{00\cdots0}_{m}$ in $\mathbb{F}_{2^m}$ if $p = 0$, and the output value is $\underbrace{00\cdots0}_{m-1}1$ in $\mathbb{F}_{2^m}$ if $p=1$.